\documentstyle[12pt]{article}
\newlength{\extraspace}
\newlength{\extraspaces}
\newcommand{\be}{\begin{equation}
\addtolength{\abovedisplayskip}{\extraspaces}
\addtolength{\belowdisplayskip}{\extraspaces}
\addtolength{\abovedisplayshortskip}{\extraspace}
\addtolength{\belowdisplayshortskip}{\extraspace}}
\newcommand{\ee}{\end{equation}}
\newcommand{\ba}{\begin{eqnarray}
\addtolength{\abovedisplayskip}{\extraspaces}
\addtolength{\belowdisplayskip}{\extraspaces}
\addtolength{\abovedisplayshortskip}{\extraspace}
\addtolength{\belowdisplayshortskip}{\extraspace}}
\newcommand{\ea}{\end{eqnarray}}
\newcommand{\nonu}{\nonumber \\[.5mm]}
\newcommand{\A}{&\!\!\!}
\begin{document}
\thispagestyle{empty}
\begin{flushright}
SIT-LP-08/03 \\
March, 2008
\end{flushright}
\vspace{7mm}
%
%
\begin{center}
{\large{\bf Systematics of linearization of $N = 2$ nonlinear SUSY \\[2mm]
in superfield formulation}} \\[20mm]
{\sc Kazunari Shima}
\footnote{
\tt e-mail: shima@sit.ac.jp} \ 
and \ 
{\sc Motomu Tsuda}
\footnote{
\tt e-mail: tsuda@sit.ac.jp} 
\\[5mm]
{\it Laboratory of Physics, 
Saitama Institute of Technology \\
Fukaya, Saitama 369-0293, Japan} \\[20mm]
\begin{abstract}
The relation between a $N = 2$ nonlinear supersymmetric (SUSY) model 
and a linear SUSY (free) theory for $N = 2$ vector supermultiplet 
accompanying the spontaneous SUSY breaking 
is systematically worked out in two-dimensional superfield formulation. 
\\[5mm]
\end{abstract}
\end{center}

\newpage

\noindent
Nonlinear (NL) realization of supersymmetry (SUSY) \cite{VA}, which induces the spontaneous SUSY breaking, 
gives the way to construct NLSUSY general relativity (GR) \cite{KS1,KS2} 
as the fundamental theory of everything in the SGM scenario 
from a compositeness viewpoint \cite{ST1,ST2}. 
The low energy physics and cosmology in NLSUSY GR are discussed \cite{ST2,ST3,STL} 
based on the linearization of NLSUSY ({\it NL-linear(L) SUSY relation}) in (Riemann-)flat spacetime. 
The linearization problem in flat spacetime was addressed mainly so far for $N = 1$ and $N = 2$ SUSY 
by studying the relation between the NLSUSY model and various LSUSY {\it free} field theories 
(with the Fayet-Iliopoulos (FI) term), for $N = 1$ scalar supermultiplet \cite{IK}-\cite{UZ}, 
for $N = 1$ ($U(1)$) (axial) vector one \cite{STT1} 
and for $N = 2$ ($SU(2) \times U(1)$) vector one \cite{STT2}. 
Linearizing $N = 3$ NLSUSY was also discussed in two dimensional spacetime ($d = 2$) \cite{ST4}. 

Recently, according to heuristic arguments, we have shown the explicit relation 
between the $N = 2$ NLSUSY model and $N = 2$ LSUSY {\it interacting} theories in $d = 2$, 
i.e. one with Yukawa interaction terms for the vector supermultiplet \cite{ST5}, 
and the other with $U(1)$ gauge interaction terms between the vector and the scalar supermultiplets 
($N = 2$ SUSY QED) \cite{ST6}. 
In order to further investigate the NL-L SUSY relation for $N \ge 2$ SUSY 
which is realistic in the SGM scenario, 
it is important to develop the systematic method of the linearization 
in superfield formulation \cite{IK,UZ} into higher $N$ SUSY theories. 
In this letter, as a preliminary to do this we discuss on the linearization of $N = 2$ NLSUSY 
($N = 2$ NL-L SUSY relation) for the $N = 2$ vector supermultiplet 
in the $d = 2$ superfield formulation at the free-theory level. 

In the linearization process, SUSY invariant relations connecting the NLSUSY model with a LSUSY theory 
are essential, where component fields in the LSUSY theory are expressed as composites 
in terms of Nambu-Goldstone (NG) fermion ({\it superon} in the SGM scenario). 
These relations are systematically obtained by defining a superfield on the following 
specific supertranslations \cite{IK,UZ} 
of superspace coordinates $(x^a, \theta^i)$ depending on the (Majorana) NG fermions $\psi^i$, 
\footnote{
Minkowski spacetime indices are denoted by $a, b, \cdots = 0, 1$ in $d = 2$
and $SO(N)$ internal indices are $i, j, \cdots = 1, 2$ for $N = 2$. 
The Minkowski spacetime metric in $d = 2$ is 
${1 \over 2}\{ \gamma^a, \gamma^b \} = \eta^{ab} = {\rm diag}(+, -)$ 
and $\sigma^{ab} = {i \over 2}[\gamma^a, \gamma^b] 
= i \epsilon^{ab} \gamma_5$ $(\epsilon^{01} = 1 = - \epsilon_{01})$, 
where we use the $\gamma$ matrices defined as $\gamma^0 = \sigma^2$, 
$\gamma^1 = i \sigma^1$, $\gamma_5 = \gamma^0 \gamma^1 = \sigma^3$ 
with $\sigma^I (I = 1, 2, 3)$ being Pauli matrices. 
}
\ba
\A \A 
x'^a = x^a + i \kappa \bar\theta^i \gamma^a \psi^i, 
\nonu
\A \A 
\theta'^i = \theta^i - \kappa \psi^i, 
\label{specific}
\ea
where $\kappa$ is a constant whose dimension is $({\rm mass})^{-1}$ in $d = 2$. 
Indeed, a superfield on $(x'^a, \theta'^i)$, 
\be
\tilde\Phi(x, \theta^i; \psi^i(x)) = \Phi(x', \theta'^i), 
\ee
transforms homogeneously as 
\be
\delta_\zeta \tilde\Phi(x, \theta^i) = \xi^a \partial_a \tilde\Phi(x, \theta^i) 
\label{tPhi-transfn}
\ee
with $\xi^a = i \kappa \bar\psi^i \gamma^a \zeta^i$, 
under superspace translations of $(x^a, \theta^i)$ 
%
%
accompanying NLSUSY transformations \cite{VA} of $\psi^i$, 
\be
\delta_\zeta \psi^i = {1 \over \kappa} \zeta^i 
- i \kappa \bar\zeta^j \gamma^a \psi^j \partial_a \psi^i, 
\label{NLSUSY}
\ee
parametrized by constant (Majorana) spinor parameters $\zeta^i$. 
The supertransformation property (\ref{tPhi-transfn}) means 
that component fields $\tilde \varphi^I(x)$ in $\tilde \Phi(x, \theta^i)$ 
do not transform each other, and SUSY invariant constraints, 
%
$\tilde \varphi^I(x) = {\rm constant}$, 
%
can be imposed, which leads to the SUSY invariant relations. 

Let us introduce a $d = 2$, $N = 2$ (general) superfield \cite{DVF,ST7} 
for the $N = 2$ vector supermultiplet, 
\ba
{\cal V}(x, \theta^i) \A = \A C(x) + \bar\theta^i \Lambda^i(x) 
+ {1 \over 2} \bar\theta^i \theta^j M^{ij}(x) 
- {1 \over 2} \bar\theta^i \theta^i M^{jj}(x) 
+ {1 \over 4} \epsilon^{ij} \bar\theta^i \gamma_5 \theta^j \phi(x) 
\nonu
\A \A 
- {i \over 4} \epsilon^{ij} \bar\theta^i \gamma_a \theta^j v^a(x) 
- {1 \over 2} \bar\theta^i \theta^i \bar\theta^j \lambda^j(x) 
- {1 \over 8} \bar\theta^i \theta^i \bar\theta^j \theta^j D(x), 
\label{VSF}
\ea
where the component fields are denoted by $(C, D)$ for two scalar fields, 
$(\Lambda^i, \lambda^i)$ for four spinor fields, 
$\phi$ for a pseudo scalar field, $v^a$ for a vector field, 
and $M^{ij} = M^{(ij)}$ $\left(= {1 \over 2}(M^{ij} + M^{ji}) \right)$ 
for three scalar fields ($M^{ii} = \delta^{ij} M^{ij}$), respectively. 
%
%
The superfield (\ref{VSF}) transforms under the superspace translations of $(x^a, \theta^i)$ as 
\be
\delta_\zeta {\cal V}(x, \theta^i) = \bar\zeta^i Q^i {\cal V}(x, \theta^i) 
\label{VSFtransfn}
\ee
with supercharges 
\be
Q_\alpha^i = {\partial \over \partial\bar\theta^{\alpha i}} 
+ i \!\!\not\!\partial \theta_\alpha^i, 
\ee
satisfying $\{ Q_\alpha^i, Q_\beta^j \} = - 2 \delta^{ij} (\gamma^a C)_{\alpha \beta} P_a$. 

The $N = 2$ superfield (\ref{VSF}) on the specific coordinates (\ref{specific}), 
\be
\tilde{\cal V}(x, \theta^i) = {\cal V}(x', \theta'^i), 
\ee
may be expanded in component fields as 
\ba
\tilde{\cal V}(x, \theta^i) \A = \A \tilde C(x) + \bar\theta^i \tilde\Lambda^i(x) 
+ {1 \over 2} \bar\theta^i \theta^j \tilde M^{ij}(x) 
- {1 \over 2} \bar\theta^i \theta^i \tilde M^{jj}(x) 
+ {1 \over 4} \epsilon^{ij} \bar\theta^i \gamma_5 \theta^j \tilde\phi(x) 
\nonu
\A \A 
- {i \over 4} \epsilon^{ij} \bar\theta^i \gamma_a \theta^j \tilde v^a(x) 
- {1 \over 2} \bar\theta^i \theta^i \bar\theta^j \tilde\lambda^j(x) 
- {1 \over 8} \bar\theta^i \theta^i \bar\theta^j \theta^j \tilde D(x), 
\label{tVSF}
\ea
where the component fields $\tilde\varphi^I(x) = (\tilde C(x), \tilde\Lambda^i(x), \tilde M^{ij}(x), \cdots)$ 
transform according to Eq.(\ref{tPhi-transfn}). 
The $\tilde\varphi^I(x)$ are evaluated 
in terms of $\varphi^I(x) = (C(x), \Lambda^i(x), M^{ij}(x) \cdots)$ in Eq.(\ref{VSF}) as follows: 
\ba
\tilde C \A = \A C' - \kappa \bar\psi^i \Lambda'^i 
+ {1 \over 2} \kappa^2 (\bar\psi^i \psi^j M'^{ij} - \bar\psi^i \psi^i M'^{jj}) 
\nonu
\A \A 
+ {1 \over 4} \kappa^2 \epsilon^{ij} \bar\psi^i \gamma_5 \psi^j \phi' 
- {i \over 4} \kappa^2 \epsilon^{ij} \bar\psi^i \gamma_a \psi^j v'^a 
+ {1 \over 2} \kappa^3 \bar\psi^i \psi^i \bar\psi^j \lambda'^j 
- {1 \over 8} \kappa^4 \bar\psi^i \psi^i \bar\psi^j \psi^j D', 
\nonu
\tilde\Lambda^i \A = \A \Lambda'^i - \kappa (\psi^j M'^{ij} - \psi^i M'^{jj}) 
- {1 \over 2} \kappa \epsilon^{ij} \gamma_5 \psi^j \phi' 
+ {i \over 2} \kappa \epsilon^{ij} \gamma_a \psi^j v'^a 
\nonu
\A \A 
- \kappa^2 \left( \psi^i \bar\psi^j \lambda'^j + {1 \over 2} \bar\psi^j \psi^j \lambda'^i \right) 
+ {1 \over 2} \kappa^3 \psi^i \bar\psi^j \psi^j D', 
\nonu
\tilde M^{ij} \A = \A M'^{ij} - \kappa \bar\psi^{(i} \lambda'^{j)} 
+ {1 \over 2} \kappa^2 \bar\psi^i \psi^j D', 
\nonu
\tilde\phi \A = \A \phi' + \kappa \epsilon^{ij} \bar\psi^i \gamma_5 \lambda'^j 
- {1 \over 2} \kappa^2 \epsilon^{ij} \bar\psi^i \gamma_5 \psi^j D', 
\nonu
\tilde v^a \A = \A v'^a + i \kappa \epsilon^{ij} \bar\psi^i \gamma^a \lambda'^j 
- {i \over 2} \kappa^2 \epsilon^{ij} \bar\psi^i \gamma^a \psi^j D', 
\nonu
\tilde\lambda^i \A = \A \lambda'^i - \kappa \psi^i D', 
\nonu
\tilde D \A = \A D', 
\label{tVSF-VSF'}
\ea
where $\varphi'^I(x) = (C'(x), \Lambda'^i(x), M'^{ij}(x) \cdots)$ are the component fields in 
\ba
{\cal V}(x', \theta'^i) \A = \A C'(x) + \bar\theta'^i \Lambda'^i(x) 
+ {1 \over 2} \bar\theta'^i \theta'^j M'^{ij}(x) 
- {1 \over 2} \bar\theta'^i \theta'^i M'^{jj}(x) 
+ {1 \over 4} \epsilon^{ij} \bar\theta'^i \gamma_5 \theta'^j \phi'(x) 
\nonu
\A \A 
- {i \over 4} \epsilon^{ij} \bar\theta'^i \gamma_a \theta'^j v^a(x) 
- {1 \over 2} \bar\theta'^i \theta'^i \bar\theta'^j \lambda'^j(x) 
- {1 \over 8} \bar\theta'^i \theta'^i \bar\theta'^j \theta'^j D(x), 
\ea
and are expanded as 
\ba
C' \A = \A C, 
\nonu
\Lambda'^i \A = \A \Lambda^i + i \kappa \!\!\not\!\partial C \psi^i, 
\nonu
M'^{ij} \A = \A M^{ij} 
- i \kappa \epsilon^{(i \vert k \vert} \epsilon^{j)l} \bar\psi^k \!\!\not\!\partial \Lambda^l 
+ {1 \over 2} \kappa^2 \epsilon^{ik} \epsilon^{jl} \bar\psi^k \psi^l \Box C, 
\nonu
\phi' \A = \A \phi + i \kappa \epsilon^{ij} \bar\psi^i \gamma_5 \!\!\not\!\partial \Lambda^j 
- {1 \over 2} \kappa^2 \epsilon^{ij} \bar\psi^i \gamma_5 \psi^j \Box C, 
\nonu
v'^a \A = \A v^a + \kappa \epsilon^{ij} \bar\psi^i \!\!\not\!\partial \gamma^a \Lambda^j 
- {i \over 2} \kappa^2 \epsilon^{ij} \bar\psi^i \gamma^a \psi^j \Box C 
+ i \kappa^2 \epsilon^{ij} \bar\psi^i \gamma^b \psi^j \partial^a \partial_b C, 
\nonu
\lambda'^i \A = \A \lambda^i + i \kappa \!\!\not\!\partial M^{ij} \psi^j 
- {i \over 2} \kappa \epsilon^{ab} \epsilon^{ij} \gamma_a \psi^j \partial_b \phi 
+ {1 \over 2} \kappa \epsilon^{ij} \left( \psi^j \partial_a v^a 
- {1 \over 2} \epsilon^{ab} \gamma_5 \psi^j F_{ab} \right) 
\nonu
\A \A
+ {1 \over 2} \kappa^2 (\Box \Lambda^i \bar\psi^j \psi^j - \Box \Lambda^j \bar\psi^i \psi^j 
- \gamma_5 \Box \Lambda^j \bar\psi^i \gamma_5 \psi^j 
- \gamma_a \Box \Lambda^j \bar\psi^i \gamma^a \psi^j 
+ 2 \!\!\not\!\partial \partial_a \Lambda^j \bar\psi^i \gamma^a \psi^j) 
\nonu
\A \A 
+ {i \over 2} \kappa^3 \!\!\not\!\partial \Box C \psi^i \bar\psi^j \psi^j, 
\nonu
D' \A = \A D + i \kappa \bar\psi^i \!\!\not\!\partial \lambda^i 
\nonu
\A \A 
- {1 \over 2} \kappa^2 \left( \bar\psi^i \psi^j \Box M^{ij} 
- {1 \over 2} \epsilon^{ij} \bar\psi^i \gamma_5 \psi^j \Box \phi 
+ {i \over 2} \epsilon^{ij} \bar\psi^i \gamma_a \psi^j \Box v^a 
- i \epsilon^{ij} \bar\psi^i \gamma_a \psi^j \partial_a \partial_b v^b \right) 
\nonu
\A \A
+ {i \over 2} \kappa^3 \bar\psi^i \psi^i \bar\psi^j \!\!\not\!\partial \Box \Lambda^j 
- {1 \over 8} \kappa^4 \bar\psi^i \psi^i \bar\psi^j \psi^j \Box^2 C. 
\label{VSF'-VSF}
\ea
Solving Eq.(\ref{tVSF-VSF'}) with respect to $\varphi^I$ in terms of ($\tilde\varphi^I$, $\psi^i$) 
and imposing SUSY (and gauge) invariant constraint on $\tilde\lambda^i$ can be considered 
as in refs.\cite{IK,UZ}, which leads to an action in terms of $\psi^i$ 
interacting with other fields in $\tilde\varphi^I$, e.g. $\tilde v^a$. 

However, focusing here on the sector which depends only on the NG fermions, 
we impose SUSY invariant constraints which eliminate the other degrees of freedom than $\psi^i$ 
as, for example, the simplest ones, 
\be
\tilde C = \tilde\Lambda^i = \tilde M^{ij} = \tilde\phi = \tilde v^a = \tilde\lambda^i = 0, 
\ \ \ \tilde D = {\xi \over \kappa} 
\label{const}
\ee
with an arbitrary dimensionless parameter $\xi$. 
Then, from Eqs.(\ref{tVSF-VSF'}) and (\ref{VSF'-VSF}) 
the relations between $\varphi^I$ and $\psi^i$ become 
\ba
C \A = \A - {1 \over 8} \xi \kappa^3 \bar\psi^i \psi^i \bar\psi^j \psi^j, 
\nonu
\Lambda^i \A = \A - {1 \over 2} \xi \kappa^2 \psi^i \bar\psi^j \psi^j 
- i \kappa \!\!\not\!\partial C \psi^i, 
\nonu
M^{ij} \A = \A {1 \over 2} \xi \kappa \bar\psi^i \psi^j 
+ i \kappa \epsilon^{(i \vert k \vert} \epsilon^{j)l} \bar\psi^k \!\!\not\!\partial \Lambda^l 
- {1 \over 2} \kappa^2 \epsilon^{ik} \epsilon^{jl} \bar\psi^k \psi^l \Box C, 
\nonu
\phi \A = \A - {1 \over 2} \xi \kappa \epsilon^{ij} \bar\psi^i \gamma_5 \psi^j 
- i \kappa \epsilon^{ij} \bar\psi^i \gamma_5 \!\!\not\!\partial \Lambda^j 
+ {1 \over 2} \kappa^2 \epsilon^{ij} \bar\psi^i \gamma_5 \psi^j \Box C, 
\nonu
v^a \A = \A - {i \over 2} \xi \kappa \epsilon^{ij} \bar\psi^i \gamma^a \psi^j 
- \kappa \epsilon^{ij} \bar\psi^i \!\!\not\!\partial \gamma^a \Lambda^j 
+ {i \over 2} \kappa^2 \epsilon^{ij} \bar\psi^i \gamma^a \psi^j \Box C 
- i \kappa^2 \epsilon^{ij} \bar\psi^i \gamma^b \psi^j \partial^a \partial_b C, 
\nonu
\lambda^i \A = \A \xi \psi^i - i \kappa \!\!\not\!\partial M^{ij} \psi^j 
+ {i \over 2} \kappa \epsilon^{ab} \epsilon^{ij} \gamma_a \psi^j \partial_b \phi 
- {1 \over 2} \kappa \epsilon^{ij} \left( \psi^j \partial_a v^a 
- {1 \over 2} \epsilon^{ab} \gamma_5 \psi^j F_{ab} \right) 
\nonu
\A \A
- {1 \over 2} \kappa^2 (\Box \Lambda^i \bar\psi^j \psi^j - \Box \Lambda^j \bar\psi^i \psi^j 
- \gamma_5 \Box \Lambda^j \bar\psi^i \gamma_5 \psi^j 
- \gamma_a \Box \Lambda^j \bar\psi^i \gamma^a \psi^j 
+ 2 \!\!\not\!\partial \partial_a \Lambda^j \bar\psi^i \gamma^a \psi^j) 
\nonu
\A \A 
- {i \over 2} \kappa^3 \!\!\not\!\partial \Box C \psi^i \bar\psi^j \psi^j, 
\nonu
D \A = \A {\xi \over \kappa} - i \kappa \bar\psi^i \!\!\not\!\partial \lambda^i 
\nonu
\A \A 
+ {1 \over 2} \kappa^2 \left( \bar\psi^i \psi^j \Box M^{ij} 
- {1 \over 2} \epsilon^{ij} \bar\psi^i \gamma_5 \psi^j \Box \phi 
+ {i \over 2} \epsilon^{ij} \bar\psi^i \gamma_a \psi^j \Box v^a 
- i \epsilon^{ij} \bar\psi^i \gamma_a \psi^j \partial_a \partial_b v^b \right) 
\nonu
\A \A
- {i \over 2} \kappa^3 \bar\psi^i \psi^i \bar\psi^j \!\!\not\!\partial \Box \Lambda^j 
+ {1 \over 8} \kappa^4 \bar\psi^i \psi^i \bar\psi^j \psi^j \Box^2 C. 
\label{VSF-VSFpsi}
\ea
We solve Eq.(\ref{VSF-VSFpsi}) entirely with respect to the component fields $\varphi^I$ 
as composites of the NG fermions $\psi^i$ and we obtain SUSY invariant relations 
for the $d = 2$, $N = 2$ vector supermultiplet in all orders of $\psi^i$ as follows: 
\ba
C \A = \A - {1 \over 8} \xi \kappa^3 \bar\psi^i \psi^i \bar\psi^j \psi^j, 
\nonu
\Lambda^i \A = \A - {1 \over 2} \xi \kappa^2 
\psi^i \bar\psi^j \psi^j (1 - i \kappa^2 \bar\psi^k \!\!\not\!\partial \psi^k), 
\nonu
M^{ij} \A = \A {1 \over 2} \xi \kappa \bar\psi^i \psi^j 
\left( 1 - i \kappa^2 \bar\psi^k \!\!\not\!\partial \psi^k 
- {1 \over 2} \kappa^4 \epsilon^{ab} \bar\psi^k \psi^l 
\partial_a \bar\psi^k \gamma_5 \partial_b \psi^l \right), 
\nonu
\phi \A = \A - {1 \over 2} \xi \kappa \epsilon^{ij} \bar\psi^i \gamma_5 \psi^j 
\left( 1 - i \kappa^2 \bar\psi^k \!\!\not\!\partial \psi^k 
- {1 \over 2} \kappa^4 \epsilon^{ab} \bar\psi^k \gamma_5 \psi^l 
\partial_a \bar\psi^k \partial_b \psi^l \right), 
\nonu
v^a \A = \A - {i \over 2} \xi \kappa \epsilon^{ij} \bar\psi^i \gamma^a \psi^j 
(1 - i \kappa^2 \bar\psi^k \!\!\not\!\partial \psi^k), 
\nonu
\lambda^i \A = \A \xi \psi^i \vert w \vert, 
\nonu
D \A = \A {\xi \over \kappa} \vert w \vert, 
\label{SUSYinv}
\ea
where $\vert w \vert$ is the determinant introduced in \cite{VA}, 
which induces a spacetime-volume differential form in the NLSUSY model, 
i.e. for the $d = 2$, $N = 2$ ($N > 2$, as well) SUSY case, 
\be
\vert w \vert = \det(w^a{}_b) = \det(\delta^a_b + t^a{}_b), 
\ \ \ t^a{}_b = - i \kappa^2 \bar\psi^i \gamma^a \partial_b \psi^i, 
\ee
expanded in terms of $t^a{}_b$ or $\psi^i$ as 
\ba
\vert w \vert \A = \A 1 + t^a{}_a + {1 \over 2!}(t^a{}_a t^b{}_b - t^a{}_b t^b{}_a) 
\nonu
\A = \A 1 - i \kappa^2 \bar\psi^i \!\!\not\!\partial \psi^i 
- {1 \over 2} \kappa^4 
(\bar\psi^i \!\!\not\!\partial \psi^i \bar\psi^j \!\!\not\!\partial \psi^j 
- \bar\psi^i \gamma^a \partial_b \psi^i \bar\psi^j \gamma^b \partial_a \psi^j) 
\nonu
\A = \A 1 - i \kappa^2 \bar\psi^i \!\!\not\!\partial \psi^i 
- {1 \over 2} \kappa^4 \epsilon^{ab} 
(\bar\psi^i \psi^j \partial_a \bar\psi^i \gamma_5 \partial_b \psi^j 
+ \bar\psi^i \gamma_5 \psi^j \partial_a \bar\psi^i \partial_b \psi^j). 
\ea
Note that all SUSY invariant relations for $\varphi^I$ in Eq.(\ref{SUSYinv}) 
are expressed as the form, 
\be
\varphi^I \sim \xi \kappa^{n-1} (\psi^i)^n \vert w \vert \ (n = 0, 1, \cdots, 4), 
\ee
where $(\psi^i)^2$ means $\bar\psi^i \psi^j$, $\epsilon^{ij} \bar\psi^i \gamma_5 \psi^j$ 
or $\epsilon^{ij} \bar\psi^i \gamma^a \psi^j$, $(\psi^i)^3 = \psi^i \bar\psi^j \psi^j$ 
and $(\psi^i)^4 = \bar\psi^i \psi^i \bar\psi^j \psi^j$. 

Let us now discuss on the relation between NLSUSY and LSUSY actions 
for the $N = 2$ vector supermultiplet in the free theory. 
The NLSUSY action \cite{VA} for $d = 2$, $N = 2$ SUSY is written in terms of $\psi^i$ as 
\ba
S_{N = 2{\rm NLSUSY}} = - {1 \over {2 \kappa^2}} \int d^2 x \ \vert w \vert, 
\label{NLSUSYaction}
\ea
which is invariant (becomes a surface term) under the NLSUSY transformations (\ref{NLSUSY}) 
due to $\delta_\zeta \vert w \vert = \partial_a (\xi^a \vert w \vert)$. 
On the other hand, the (free) action for the $N = 2$ vector supermultiplet with the FI $D$ term 
is given by using the superfield (\ref{VSF}) as follows: 
\be
S_{{\cal V}0} = \int d^2 x \left[ \int d^2 \theta^i {\cal L}_0 (x, \theta^i) 
+ \int d^4 \theta^i {\cal L}_{\rm FI} (x, \theta^i) \right]_{\theta^i = 0}, 
\label{V0}
\ee
where 
\ba
\A \A 
{\cal L}_0 (x, \theta^i) = {1 \over 32} (\overline{D^j {\cal W}^{kl}} D^j {\cal W}^{kl} 
+ \overline{D^j {\cal W}_5^{kl}} D^j {\cal W}_5^{kl}), 
\\
\A \A 
{\cal L}_{\rm FI} (x, \theta^i) = {\xi \over {2 \kappa}} {\cal V} 
\ea
with 
\ba
\A \A 
D_\alpha^i = {\partial \over \partial\bar\theta^{\alpha i}} 
- i \!\!\not\!\partial \theta_\alpha^i, 
\\
\A \A 
{\cal W}^{ij} = \bar D^i D^j {\cal V}, 
\ \ \ {\cal W}_5^{ij} = \bar D^i \gamma_5 D^j {\cal V}. 
\ea
The action (\ref{V0}) in the WZ gauge gives the $N = 2$ LSUSY (free) action 
for the minimal off-shell component fields $(A, \phi, v^a, \lambda^i, D)$ 
with $A = M^{ii} (= M^{11} + M^{22})$ \cite{ST7}, 
\ba
\A \A 
S_{V0} = S_{{\cal V}0} \ \vert_{\rm WZ\ gauge} 
\nonu
\A \A 
= \int d^2 x \left\{ - {1 \over 4} (F_{ab})^2 
+ {i \over 2} \bar\lambda^i \!\!\not\!\partial \lambda^i 
+ {1 \over 2} (\partial_a A)^2 + {1 \over 2} (\partial_a \phi)^2 
+ {1 \over 2} D^2 - {\xi \over \kappa} D 
\right\}, 
\label{V0min}
\ea
where the field equation for the auxiliary field, $D = {\xi \over \kappa}$, 
indicates the spontaneous SUSY breaking. 

The relation between the actions (\ref{NLSUSYaction}) and (\ref{V0min}) with $\xi^2 = 1$, i.e. 
\be
S_{N = 2{\rm NLSUSY}} = S_{V0} + [{\rm suface\ terms}], 
\ee
can be shown by substituting SUSY invariant relations for the minimal off-shell vector supermultiplet, 
$(A, \phi, v^a, \lambda^i, D)(\psi^i)$, into the action (\ref{V0min}) directly \cite{ST5}. 
Here let us show that the LSUSY action (\ref{V0}) exactly reduces to the NLSUSY action (\ref{NLSUSYaction}) 
when $\xi^2 = 1$ by using the superfield (\ref{tVSF}) in the SUSY invariant constraints (\ref{const}), 
\be
\tilde{\cal V}(x, \theta^i) 
= - {\xi \over {8 \kappa}} \bar\theta^i \theta^i \bar\theta^j \theta^j, 
\label{VSF-const}
\ee
which lead to the SUSY invariant relations (\ref{SUSYinv}): 
Indeed, by changing the integration variables in Eq.(\ref{V0}) from $(x, \theta^i)$ to $(x', \theta'^i)$, 
we obtain 
\ba
S_{{\cal V}0} \A = \A \int d^2 x' \left[ \int d^2 \theta'^i {\cal L}_0 (x', \theta'^i) 
+ \int d^4 \theta'^i {\cal L}_{\rm FI} (x', \theta'^i) \right]_{\theta'^i = 0} 
\nonu
\A = \A \int d^2 x \left[ \int d^2 \theta^i J(x, \theta^i) \tilde{\cal L}_0 (x, \theta^i) 
+ \int d^4 \theta^i J(x, \theta^i) \tilde{\cal L}_{\rm FI} (x, \theta^i) \right]_{\theta^i = 0}, 
\label{V0change}
\ea
where 
\ba
\A \A 
\tilde{\cal L}_0 (x, \theta^i) = {1 \over 32} (\overline{D'^j \tilde{\cal W}^{kl}} D'^j \tilde{\cal W}^{kl} 
+ \overline{D'^j \tilde{\cal W}_5^{kl}} D'^j \tilde{\cal W}_5^{kl}), 
\\
\A \A 
\tilde{\cal L}_{\rm FI} (x, \theta^i) = {\xi \over {2 \kappa}} \tilde{\cal V} 
\ea
with 
\ba
\A \A 
D_\alpha'^i = {\partial \over \partial\bar\theta'^{\alpha i}} 
- i \!\!\not\!\partial' \theta_\alpha'^i, 
\label{diff'}
\\
\A \A 
\tilde{\cal W}^{ij} = \bar D'^i D'^j \tilde{\cal V}, 
\ \ \ \tilde{\cal W}_5^{ij} = \bar D'^i \gamma_5 D'^j \tilde{\cal V}. 
\label{tW-tW5}
\ea
In Eq.(\ref{V0change}) the $J(x, \theta^i)$ means the Jacobian given by 
\be
J(x, \theta^i) = {\rm sdet} M 
= \vert w \vert \det(\delta_b^a - i \kappa \nabla_b \bar\psi^i \gamma^a \theta^i), 
\label{Jacob}
\ee
where sdet is the superdeterminant, the supermatix $M$ and the ``covariant'' derivative $\nabla_a$ \cite{UZ} 
are defined by 
\ba
\A \A 
M = {{\partial(x', \theta'^i)} \over {\partial(x, \theta^j)}} 
= \left( 
\begin{array}{ccc}
\delta_b^a - i \kappa \partial_b \bar\psi^i \gamma^a \theta^i & - \kappa \partial_b \bar\psi^i \\
i \kappa \gamma^a \psi^j & \delta^{ij} 
\end{array}
\right), 
\nonu
\A \A 
\nabla_a = (w^{-1})_a{}^b \partial_b. 
\ea
Also, the transformation of derivatives is 
\be
\left( 
\begin{array}{ccc}
\partial'_a \\
{\partial \over \partial\bar\theta'^i} 
\end{array}
\right)
= M^{-1} 
\left( 
\begin{array}{ccc}
\partial_a \\
{\partial \over \partial\bar\theta^i} 
\end{array}
\right), 
\ \ \ 
M^{-1} = \left( 
\begin{array}{ccc}
v_a{}^b & \kappa v_a{}^b \partial_b \bar\psi^j \\
- i \kappa \gamma^a \psi^i v_a{}^b 
& \delta^{ij} - i \kappa^2 \gamma^a \psi^i \partial_b \bar\psi^j v_a{}^b 
\end{array}
\right), 
\ee
where $v_a{}^b$ is determined from 
\be
w_a{}^c (\delta_c^d - i \kappa \nabla_c \bar\psi^i \gamma^d \theta^i) \ v_d{}^b = \delta_a^b, 
\ee
and is solved as 
\be
v_a{}^b = (w^{-1})_a{}^b + i \kappa \nabla_a \bar\psi^i \gamma^c \theta^i (w^{-1})_c{}^b 
+ {\cal O}((\theta^i)^2). 
\ee
Then the differential operators (\ref{diff'}) are expressed by means of 
$\left( \partial_a, {\partial \over \partial\bar\theta^i} \right)$ as 
\be
D_\alpha'^i = {\partial \over \partial\bar\theta^{\alpha i}} 
- i \gamma^a \theta_\alpha^i v_a{}^b 
\left( \partial_b 
+ \kappa \partial_b \bar\psi^j {\partial \over \partial\bar\theta^{\alpha j}} \right), 
\label{diff}
\ee
By substituting Eqs.(\ref{VSF-const}), (\ref{Jacob}) and (\ref{diff}) into Eq.(\ref{V0change}), 
the relation between the actions (\ref{NLSUSYaction}) and (\ref{V0}), 
\be
S_{N = 2{\rm NLSUSY}} = S_{{\cal V}0}, 
\label{NLSUSY-LSUSY}
\ee
is shown when $\xi^2 = 1$. 


We summarize our results as follows. 
In this letter we have systematically linearized $N = 2$ NLSUSY 
in the $d = 2$ superfield formulation for the $N = 2$ vector supermultiplet. 
Based on the $d = 2$, $N = 2$ superfield (\ref{VSF}) 
the relation between the component fields $\tilde \varphi^I(x)$ in Eq.(\ref{tVSF}) 
and $\varphi^I(x)$ in Eq.(\ref{VSF}) are given as in Eqs.(\ref{tVSF-VSF'}) and (\ref{VSF'-VSF}). 
By imposing the (simplest) SUSY invariant constraints (\ref{const}), we have obtained the SUSY invariant 
relations (\ref{SUSYinv}) uniquely, which coincide with those for the minimal off-shell 
vector supermultiplet obtained heuristicly in Ref.\cite{ST5}. 
The $N = 2$ NLSUSY action (\ref{NLSUSYaction}) is just reproduced when $\xi^2 = 1$ 
by substituting the SUSY invariant relations into the $N = 2$ LSUSY free action 
with the FI $D$ term (\ref{V0}), i.e. we have shown the relation (\ref{NLSUSY-LSUSY}) 
in the free theory from the superfield formulation. 
The extensions of the superfield method for the linearization to higher $N$ NLSUSY and to $d = 4$ are important. 
The Yukawa interaction terms \cite{ST5} and the coupling of matter supermultiplets (SUSY QED) \cite{ST6} 
in the linearization framework of this letter are interesting problems under the investigation.

\newpage

%
\newcommand{\NP}[1]{{\it Nucl.\ Phys.\ }{\bf #1}}
\newcommand{\PL}[1]{{\it Phys.\ Lett.\ }{\bf #1}}
\newcommand{\CMP}[1]{{\it Commun.\ Math.\ Phys.\ }{\bf #1}}
\newcommand{\MPL}[1]{{\it Mod.\ Phys.\ Lett.\ }{\bf #1}}
\newcommand{\IJMP}[1]{{\it Int.\ J. Mod.\ Phys.\ }{\bf #1}}
\newcommand{\PR}[1]{{\it Phys.\ Rev.\ }{\bf #1}}
\newcommand{\PRL}[1]{{\it Phys.\ Rev.\ Lett.\ }{\bf #1}}
\newcommand{\PTP}[1]{{\it Prog.\ Theor.\ Phys.\ }{\bf #1}}
\newcommand{\PTPS}[1]{{\it Prog.\ Theor.\ Phys.\ Suppl.\ }{\bf #1}}
\newcommand{\AP}[1]{{\it Ann.\ Phys.\ }{\bf #1}}

\end{document}